\documentclass[prd,aps,showpacs,tightenlines]{revtex4}  
\usepackage{mathrsfs}
\usepackage{amsmath}
\usepackage{amssymb}
\usepackage{epsfig}
\usepackage{graphicx}
\usepackage{booktabs}
\usepackage{multirow}
\usepackage{subfigure}
\usepackage{color}
\begin{document}
\newcommand{\psl}{ p \hspace{-1.8truemm}/ }
\newcommand{\nsl}{ n \hspace{-2.2truemm}/ }
\newcommand{\vsl}{ v \hspace{-2.2truemm}/ }
\newcommand{\epsl}{\epsilon \hspace{-1.8truemm}/\,  }

\title{Branching ratios and $CP$ asymmetries of $B\rightarrow \chi_{c1}K(\pi)  $ decays }
\author{Zhou Rui$^1$}\email{jindui1127@126.com}
\author{Qiang Zhao$^1$}
\author{Li-li Zhang$^2$}
\affiliation{$^1$College of Sciences, North China University of Science and Technology,
                          Tangshan 063009,  China}
\affiliation{$^2$Center for Publishing, North China University of Science and Technology,
                          Tangshan 063009,  China}
\date{\today}
\begin{abstract}
We investigate the exclusive nonleptonic decays  $B\rightarrow \chi_{c1}K(\pi)$  in the conventional perturbative QCD (PQCD) formalism.
The predictions of branching ratios  and   $CP$  asymmetries are given in detail.
We compare our results with available experimental data as well as predictions of other theoretical studies existing
in the literature.
It seems that   the branching ratios of $B\rightarrow \chi_{c1} K$
are more consistent with data than the earlier analyses.
For the Cabibbo-suppressed $B_s$ decays, the branching ratios can reach the order of $10^{-5}$,
 which would be straight forward for experimental observations.
 The numerical results show that the direct $CP$ asymmetries of the concerned decays are rather small.
 The mixing-induced $CP$ asymmetry in the $B^0\rightarrow \chi_{c1}K_S$ is very close to $
 \sin{2\beta}$, which suggests that this channel
 offer an alternative method for measuring
the Cabbibo-Kobayashi-Maskawa (CKM) angle $\beta$.
 The obtained results in the present work could be tested by further experiments
in the LHCb and forthcoming Belle II.

\end{abstract}



\maketitle

\section{Introduction}
Decays of $B$ mesons to two-body final states including a charmonium meson,
proceed via a $b\rightarrow s c\bar{c}$ or $b\rightarrow d c\bar{c}$ quark transition,
provide us with a playground for understanding the features of
$CP$ violation in the $B$ meson system.
For the Cabibbo-favored $b\rightarrow s c\bar{c}$ modes, such as $B\rightarrow J/\psi K$,
the tree and  penguin contributions   have the same weak phase to order $\lambda^2$
and thus no direct $CP$ violation is expected.
In the Cabibbo-suppressed $b\rightarrow d c\bar{c}$ transitions, however,
the tree and penguin contributions have different phases and the $CP$ asymmetries may appear at the percent level,
 e.g. $B\rightarrow J/\psi \pi$. Any asymmetry larger than this magnitude would be the harbingers of new physics and  of significant interest.
The current experimental measurements of the direct $CP$ violations for exclusive decays
of $B$ mesons to charmonium and kaon or pion final states, which have been averaged
by the Particle Data Group (PDG) \cite{pdg2016}, are summarized below:
\begin{eqnarray}\label{eq:dirchi1}
A_{J/\psi K^+} &=& 0.003\pm 0.006, \quad A_{\psi(2S) K^+} = 0.012\pm 0.020, \quad  A_{\chi_{c1} K^+} = -0.009\pm 0.033, \nonumber\\
A_{J/\psi \pi^+} &=& 0.001\pm 0.028, \quad A_{\psi(2S) \pi^+} = 0.03\pm 0.06, \quad  A_{\chi_{c1} \pi^+} = 0.07\pm 0.18,
\end{eqnarray}
where statistical and systematic uncertainties have been added in quadrature.
The neutral $B$ decays to $CP$ eigenstates containing a charmonium and $K_S^0$ 
are  regarded as the golden mode for extracting the mixing-induced
$CP$ asymmetry parameter  $S_f=-\eta \sin{2\beta}$, $\beta$ being
the weak phase of the CKM matrix element $V_{td}$
 and $\eta$ is the $CP$ eigenvalue of the final state $f$.
 The latest average of
 Heavy Flavor Averaging Group (HFAVG)  \cite{HFLAV2016}
  gave $S_{J/\psi K^0_S}=0.665 \pm 0.024 $,  $S_{\psi(2S) K^0_S}=0.807\pm 0.067$, and $S_{\chi_{c1} K^0_S}=0.632\pm 0.099$
 corresponding to the $\eta=-1$ modes  $J/\psi K^0_S$, $\psi(2S) K^0_S$, and $\chi_{c1} K^0_S$, respectively.
The small spread in the  $CP$ asymmetry parameters between different charmonia may indicate
the penguin contributions in these decays are split according to 
  different $c\bar{c}$ systems.

On the other hand,  the dominant mechanism for charmonium production in the $B$ meson decay is
color-suppressed, so precise measurements of rates to the exclusive modes can provide
valuable insight into the dynamics of strong interactions in heavy meson systems.
In particular, any mode involving various excitations of the $c\bar{c}$  assignments such
as $P$-wave charmonium  productions could be an alternative to that for $S$-wave counterparts, 
and they could give additional and complementary information about the exclusive charmonium  decays of the $B$ meson.
Under the factorization hypothesis, those decays are allowed when the charm-anticharm  pair hadronizes to $\chi_{c1}$,
but suppressed when the quark  pair hadronizes to $\chi_{c0,c2}$ and $h_c$ due to the spin-parity and vector current conservation \cite{prd66037503}. Of course it is possible if there is an exchange of an additional  gluon, which is the so-called  non-factorizable contributions.
 Therefore, these modes are particularly illuminating as they
provide valuable information for understanding  of   the non-factorizable mechanism.
In fact, both the $BABAR$ \cite{prd78091101,prl102132001} and Belle \cite{prl88031802,prl107091803} have found a surprisingly large branching ratios of $B\rightarrow \chi_{c0,c2}K^{(*)}$ decays, which are even comparable to that of the factorization-allowed decay, such as $B\rightarrow \chi_{c1}K$.
Besides,   some other $P$-wave charmonium productions in  $B$ meson decay have been observed by several experimental collaborations, such as
$B\rightarrow \chi_{c1} \pi$ \cite{prd74051103,prd78091104},
$B\rightarrow \chi_{c1} K^{(*)}$ \cite{prl89011803,prl94141801,plb634155},
$B\rightarrow h_{c} K^{(*)}$ \cite{prd74012007,prd78012006}, and so on. 
Most recently, the Belle collaboration present the measurement of the absolute branching fractions of $B^+\rightarrow X_{cc} K^+$,
where $X_{cc}$ denotes nine charmonium states: $\eta_c$, $J/\psi$, $\chi_{c0}$, $\chi_{c1}$, $\eta_c(2S)$, $\psi(2S)$, $X(3870)$, $X(3872)$, and $X(3915)$ \cite{prd97012005}.
As for hadronic $B_s$ decays, the first observation of the decay  $B_{s}\rightarrow \chi_{c1}\phi$ \cite{npb874663} are reported
by the LHCb experiment, meanwhile, some relative ratios of the branching ratios for $B$ meson decays into $\chi_{c1}$ and $J/\psi$ mesons are also measured, which would discriminate the mass dependence from the quantum number dependence \cite{prd76031102}.

Phenomenologically the $B$ meson decays into various $P$-wave charmonium
 have been studied in different approaches.
In Ref. \cite{plb59191}, the authors  analyze the soft nonfactorizable contributions to
$B\rightarrow (\eta_c,J/\psi,\chi_{c0,c1})K$ decays  by using the light-cone sum rules (LCSR) approach, and they found
the nonfactorizable contributions are sizable for $B\rightarrow (\chi_{c1},J/\psi) K$, while for
the $B$ decays into a (pseudo) scalar charmonia, the nonfactorizable   contributions are too small to
accommodate the data.
In Ref. \cite{prd71114008}, the same decay modes are studied by using  a hybrid PQCD approach, in which
the factorizable contributions are   treated in naive factorization (FA).
The nonfactorizable diagrams are evaluated utilizing the conventional PQCD formalism, which is free
from the endpoint singularities.
Within the framework of QCD factorization (QCDF) \cite{qcdf},
the exclusive $B$ decays to $P$-wave charmonium states were discussed   earlier
 \cite{plb568127,prd69054009,plb619313,ctp48885,0607221,npb811155,prd87074035},
 and they found the soft contributions may be large since there exist infrared divergences
 in the vertex corrections and end-point singularities in the leading twist spectator corrections.
Subsequently, the explicit calculations in Refs. \cite{npb811155,prd87074035} show that the
infrared divergences arising from  vertex corrections cancel in the $B\rightarrow \chi_{c1} K$
 decay as in the case of $B\rightarrow J/\psi K$.

Based on  the $k_T$ factorization theorem, after including the parton transverse momentum $k_T$ and
threshold resummations,
both factorizable and  nonfactorizable  decay amplitudes are  calculable without  endpoint singularity.
For detailed discussions of this approach, one can consult Refs. \cite{pqcd1,pqcd2}.
In general, the PQCD approach is suitable for describing various charmonium decays of $B$ meson
\cite{prd74114029,jhep03009,cpc34937,cpc341680,prd86011501,prd89094010,plb772719} and has a good predictive power.
In our previous work \cite{epjc77610}, the PQCD approach had been applied to study the $B\rightarrow J/\psi V, \psi(2S) V$ decays with $V$ encompasses $\rho, \omega, K^{*}, \phi$ and gave satisfactory results.
The main focus of this work lies on the $B\rightarrow \chi_{c1} K(\pi)$ decays,
while other 
factorization forbidden decays  are beyond the scope of the present analysis
because of the appearance of nonvanishing infrared divergences arising from nonfactorizable vertex corrections.
As mentioned above, the $B\rightarrow \chi_{c1}K$ decay had been analyzed in a hybrid  PQCD approach \cite{prd71114008},
where the factorizable contributions were parameterized in FA with the $B\rightarrow K$ form factors taken from the light-front QCD \cite{prd69074025}. Here both the factorizable   and nonfactorizable contributions 
are evaluated utilizing the conventional PQCD formalism.
This is  the main difference between \cite{prd71114008} and  our calculations.
Besides, we update the $\chi_{c1}$ distribution amplitudes (DAs) according to our recent work \cite{171208928},
 where the new universal nonperturbative objects are successful  in describing various $P$-wave charmonium productions in
 the case of $B_c$ meson decays.
Thus it is motivated to check for validity of 
the same scenario in the $B$ meson  decays.
For the vertex corrections,
we employ the most recent updated results   from the QCDF \cite{npb811155,prd87074035}.
 Finally, we also investigate the $CP$   asymmetry parameters including
 the Cabibbo-suppressed    $B\rightarrow \chi_{c1} \pi$ and  $B_s\rightarrow \chi_{c1} \bar{K}$ decays,
 which may be tested by the LHCb and Belle-II with continuously increasing high precision measurements.

 The presentation of the paper is as follows.
 After this introduction, we formulate the decay amplitudes of $B\rightarrow\chi_{c1}K(\pi)$ in the PQCD approach.
 In Sec. \ref{sec:results}, we give the numerical results and discussions.
Finally, we conclude in Sec. \ref{sec:sum} with a summary.

\section{Theoretical details}\label{sec:framework}

The effective Hamiltonian relevant for $B\rightarrow \chi_{c1} K(\pi)$ has  the following form \cite{rmp681125}:
\begin{eqnarray}\label{eq:operator}
\mathcal{H}_{eff}&=&\frac{G_F}{\sqrt{2}}\{\xi_c[C_1(\mu)(\bar{q}_ic_j)_{V-A}(\bar{c}_jb_i)_{V-A}
+C_2(\mu)(\bar{q}_ic_i)_{V-A}(\bar{c}_jb_j)_{V-A}]\nonumber\\&&
-\xi_t[C_3(\mu)(\bar{q}_ib_i)_{V-A}(\bar{q}'_jq'_j)_{V-A}+C_4(\mu)(\bar{q}_ib_j)_{V-A}(\bar{q}'_jq'_i)_{V-A}\nonumber\\&&
+C_5(\mu)(\bar{q}_ib_i)_{V-A}(\bar{q}'_jq'_j)_{V+A}+C_6(\mu)(\bar{q}_ib_j)_{V-A}(\bar{q}'_jq'_i)_{V+A} \nonumber\\&&
+C_7(\mu)\frac{3}{2}(\bar{q}_ib_i)_{V-A} \sum_{q'}e_{q'}(\bar{q}'_jq'_j)_{V+A}+C_8(\mu)\frac{3}{2}(\bar{q}_ib_j)_{V-A} \sum_{q'}e_{q'}(\bar{q}'_jq'_i)_{V+A}\nonumber\\&&
+C_9(\mu)\frac{3}{2}(\bar{q}_ib_i)_{V-A} \sum_{q'}e_{q'}(\bar{q}'_jq'_j)_{V-A}+C_{10}(\mu)\frac{3}{2}(\bar{q}_ib_j)_{V-A} \sum_{q'}e_{q'}(\bar{q}'_jq'_i)_{V-A}]\},
\end{eqnarray}
where $V\pm A\equiv \gamma_{\mu}(1\pm\gamma_5)$, $i,j$ are colour indices, $e_{q'}$ are the electric charges of
the quarks in units of $|e|$, and a summation over $q'=u,d,s,c,b$ is implied.
$G_F$ is the Fermi constant and  $\xi_{c(t)}=V^*_{c(t)b}V_{c(t)q}$ with $q=d,s$ are the products of CKM matrix element.
 $C_i(\mu)$ are the QCD corrected Wilson coefficients at the renormalization scale $\mu$.

\begin{figure}[tbp]
\centerline{\epsfxsize=7cm \epsffile{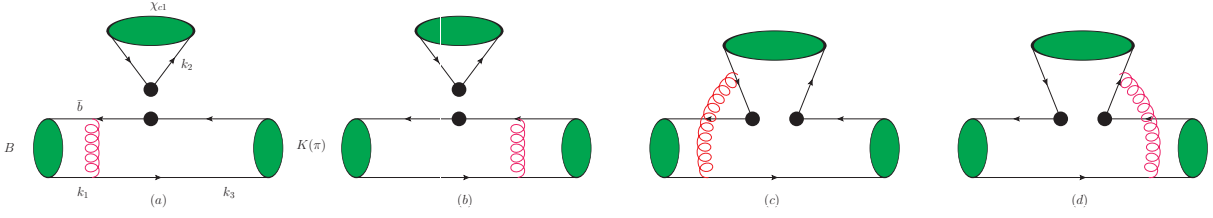}}
\vspace{1cm}
\caption{The typical leading-order Feynman diagrams for the decays  $B_c \to \chi _{c1} K(\pi)$.
(a,b) The factorizable  diagrams,  and (c,d) the nonfactorizable diagrams.}
\label{fig:femy}
\end{figure}

In the PQCD framework, the decay amplitude is factorized
into the convolution of the meson wave functions, the hard
scattering kernels and the Wilson coefficients, which stand for 
the dynamics  below, around, and above the $b$ quark mass, respectively. The formalism  can be written as
\begin{eqnarray}
\label{eq:ampu}
\mathcal{A}(B\rightarrow \chi_{c1} K(\pi)) = \int d^4k_1d^4k_2d^4k_3 Tr[C(t)\Phi_B(k_1)\Phi_{\chi_{c1}}(k_2)\Psi_{K(\pi)}(k_3)
H(k_1,k_2,k_3,t)],
\end{eqnarray}
where $k_i$ are  the quark momentum in each meson, and ``Tr'' denotes
the trace over all Dirac structures and color indices.
$C(t)$ is the standard perturbative QCD coefficient, 
which evolve from the $W$ boson mass down to the renormalization scale $t$.
The meson wave functions $\Phi$ absorb the nonperturbative dynamics in the hadronization processes.
The explicit expression of $\Phi_{\chi_{c1}}$
refer to our  previous work \cite{171208928}, while for  $\Phi_{B}$ and  $\Phi_{K(\pi)}$
 including some relevant parameters are the same as those used in \cite{prd76074018}.
The remaining finite contribution is assigned to a hard amplitude $H$,
which contains  the  four quark operator and a hard gluon connecting the spectator quark.
This renders the perturbative calculations in an effective six quark interaction form.
The relevant Feynman diagrams are shown in Fig .\ref{fig:femy}.
Below we present the calculation of the hard amplitude in the PQCD approach.

The calculation is carried out in the rest frame of $B$ meson.
The $B$ meson momentum $P_1$, the $\chi_{c1}$ meson momentum $P_2$, the light meson momentum $P_3$, and
the quark momenta $k_1$, $k_2$, and $k_3$ corresponding to  $B$, $\chi_{c1}$, and light mesons, respectively,
whose notation are displayed in Fig \ref{fig:femy}(a), are chosen as
\begin{eqnarray}
 P_1&=&\frac{M}{\sqrt{2}}(1,1,\textbf{0}_{\rm T}),\quad P_2=\frac{M}{\sqrt{2}}(1,r^2,\textbf{0}_{\rm T}),\quad  P_3=\frac{M}{\sqrt{2}}(0,1-r^2,\textbf{0}_{\rm T}),\nonumber\\
  k_1&=&(\frac{M}{\sqrt{2}}x_1,0,\textbf{k}_{\rm 1T}),\quad k_2=(\frac{M}{\sqrt{2}}x_2,\frac{M}{\sqrt{2}}x_2r^2,\textbf{k}_{2\rm T}),\quad  k_3=(0,\frac{M}{\sqrt{2}}x_3(1-r^2),\textbf{k}_{\rm 3T}),
\end{eqnarray}
with the mass ratio $r=m_{\chi_{c1}}/M$ and $m_{\chi_{c1}},M$ are the masses of the charmonium and $B$ meson, respectively.
The $k_{iT}$, $x_i$ represent the transverse momentum and longitudinal
momentum fraction of the quark inside the meson.
Like vector  mesons, axial-vector charmonium $\chi_{c1}$ also carry spin degrees of freedom.
For the decays under study,  only longitudinally polarized $\chi_{c1}$ produced with the   polarization vector
$\epsilon_{2L}=\frac{1}{\sqrt{2}r}(1,-r^2,\textbf{0}_{\rm T})$,
which satisfy the normalization $(\epsilon_{2L})^2 =-1$  and
the orthogonality $\epsilon_{2L} \cdot P_2=0$.
According to Eq. (\ref{eq:ampu}), the full decay amplitudes for the considered decays is written as
\begin{eqnarray}\label{eq:alnt}
\mathcal{A}&=&\xi_c
\Big [a_2\mathcal{F}^{LL} +C_2\mathcal{M}^{LL} \Big]
-\xi_t\Big [(a_3+a_9)\mathcal{F}^{LL}+  
(a_5+a_7)\mathcal{F}^{LR}
+(C_4+C_{10})\mathcal{M}^{LL}+(C_6+C_8)\mathcal{M}^{SP}\Big ].
\end{eqnarray}
The QCD factors $a_i$ appear in decay amplitudes, which encode dynamics of the decay,
are related to $C_i$ as follows:
 \begin{eqnarray}
a_2=C_1+\frac{1}{3}C_2, \quad  a_{i}=C_i+\frac{1}{3}C_{i+1} \quad \text{for} \quad i=3,5,7,9.
\end{eqnarray}
The superscript $LL$, $LR$, and $SP$  refers to the contributions from $(V-A)\otimes(V-A)$, $(V-A)\otimes(V+A)$
and $(S-P)\otimes(S+P)$ operators, respectively.
$\mathcal{F}(\mathcal{M})$ describes the contributions from the factorizable (nonfactorizable) diagrams in Fig. \ref{fig:femy},
which can be expressed as
\begin{eqnarray}\label{eq:flll}
\mathcal{F}^{LL}&=&8\pi C_f f_{\chi_{c1}} M^4\int_0^1dx_1dx_3\int_0^{\infty}b_1b_3db_1db_3 \phi_B(x_1,b_1)\nonumber\\&&\{
[(r^2-1)\phi^A(x_3)((r^2-1)x_3-1)+(r^2-1)\phi^P(x_3)r_p(2x_3-1)-\nonumber\\&&
\phi^T(x_3)r_p(2x_3-1-r^2(1+2x_3))]\alpha_s(t_a)S_{ab}(t_a)h(\alpha_e,\beta_a,b_1,b_3)S_t(x_1)\nonumber\\&&
-2r_{p}(1-r^2)\phi^P(x_3)\alpha_s(t_b)S_{ab}(t_b)h(\alpha_e,\beta_b,b_1,b_3)S_t(x_3)
\},
\end{eqnarray}
\begin{eqnarray}\label{eq:mlll}
\mathcal{M}^{LL}&=&-16\sqrt{\frac{2}{3}}\pi C_f M^4\int_0^1dx_1dx_2dx_3\int_0^{\infty}b_1b_2db_1db_2 \phi_B(x_1,b_1)\nonumber\\&&
[(r^2-1)\phi^A(x_3)+2r_p\phi^T(x_3)][\psi^L(x_2)(r^2(x_1+x_3-2x_2)-x_3)+2rr_c\psi^t(x_2)]
\nonumber\\&&\alpha_s(t_d)S_{cd}(t_d)h(\alpha_e,\beta_d,b_1,b_2),
\end{eqnarray}
\begin{eqnarray}
\mathcal{F}^{LR}=-\mathcal{F}^{LL},\quad  \mathcal{M}^{SP}=\mathcal{M}^{LL},
\end{eqnarray}
with $r_c=m_c/M$ and $m_c$ is the charm quark mass;  $C_f=4/3$ is a color factor;
 $f_{\chi_{c1}}$ is the vector decay constant of the $\chi_{c1}$ meson.
The hard scales $t$,  functions $h$, Sudakov factors $S(t)$, and the threshold resummation factor $S_t(x)$
 refer to Appendix A of Ref. \cite{epjc77610} for details.
 Note that the coefficient $-\frac{1}{\sqrt{2}}$ appears for $B\rightarrow \chi_{c1}\pi^0$ mode
 due to the $\pi^0$ meson generated from a pure $d\bar{d}$ source.
 In addition, we also consider the  vertex corrections to the factorizable diagrams in Fig. \ref{fig:femy}.
 As stated in Refs \cite{npb811155,prd87074035}, the infrared divergences cancel when one sums over all the  vertex corrections
 for the $B\rightarrow\chi_{c1}K$ decay,  just like the decays to $S-$wave charmonia,
 where the same hard vertex corrections are infrared finite.
 Therefore, it is not necessary  to introduce  the quark  transverse momentum $k_T$ at the end-point region \cite{prd72114005}.
 The calculations are then performed  in the collinear factorization theorem instead of the $k_T$ factorization theorem.
As a consequence,  we can simply quote the QCDF expressions for the vertex corrections.
According to the power counting of QCDF \cite{qcdf}, the hard spectator (nonfactorizable) interaction is of the same
order in $\alpha_s$ as the vertex corrections in the heavy-quark limit.
So the  corrections to the nonfactorizable diagrams Fig. \ref{fig:femy} (c) and (d) are further  power suppressed. 
In fact, the full next-to-leading-order (NLO) corrections to the charmonium $B$ decays
under the PQCD framework are still  unavailable, 
but the significant vertex corrections should be included in this work when comparing with the predictions of the QCDF.
As usual, the vertex corrections  effects 
can be combined into the   coefficients $a_i$ in Eq. (\ref{eq:alnt}) as  \cite{qcdf}
\begin{eqnarray}\label{eq:vertex}
a_2&\rightarrow & a_2 +\frac{\alpha_s}{4\pi}\frac{C_f}{N_c}C_2
 \left[-18-12\text{ln}(\frac{t}{m_b})+f_I\right ] ,\nonumber\\ a_3+a_9 &\rightarrow &a_3+a_9
+\frac{\alpha_s}{4\pi}\frac{C_f}{N_c}(C_4+C_{10}) \left[-18-12\text{ln}(\frac{t}{m_b})+f_I\right ],\nonumber\\
a_5+a_7  &\rightarrow &a_5+a_7-\frac{\alpha_s}{4\pi}\frac{C_f}{N_c}(C_6+C_{8}) \left[-6-12\text{ln}(\frac{t}{m_b})+f_I\right ],
\end{eqnarray}
where  $N_c$ is the color factor.
The quantity $f_I$   account for vertex corrections, whose detail calculations
 can be found in Refs. \cite{prd87074035,npb811155}.
\section{Numerical results and discussions}\label{sec:results}
To proceed the numerical analysis, it is useful to summarize all of the input quantities entering the PQCD approach below:
\begin{itemize}
\item[$\bullet$] For the  masses (in GeV) \cite{pdg2016}: $M_B=5.28$, \quad $M_{B_s}=5.37$,
\quad $m_{\chi_{c1}}=3.511$,\quad $m_b(\text{pole})=4.8$, \quad $\bar{m}_c(\bar{m}_c)=1.275$.
\item[$\bullet$] For the lifetimes (in ps) \cite{pdg2016}: $\tau_{B_s}=1.51, \quad \tau_{B_0}=1.52,\quad \tau_{B^+}=1.638$.
\item[$\bullet$] For the Wolfenstein parameters   \cite{pdg2016}:
$\lambda = 0.22506$,\quad $A=0.811$, \quad $\bar{\rho}=0.124$,\quad $\bar{\eta}=0.356$.
\item[$\bullet$] For the Gegenbauer moments at the scale of $\mu=1$ GeV
 \cite{prd76074018}: $a_1^K=0.17$, \quad $a_2^K=0.2$, \quad  $a_1^{\pi}=0$, \quad $a_2^{\pi}=0.44$.
\item[$\bullet$] For the  decay constants (in GeV): $f_B=0.19$ \cite{pdg2016}, \quad $f_{B_s}=0.227$ \cite{pdg2016},
\quad $f_{\chi_{c1}}=0.335$ \cite{prd71114008}, \quad $f_{\pi}=0.131$ \cite{prd76074018}, \quad $f_{K}=0.16$ \cite{prd76074018}.
\end{itemize}
The chiral factor $m_0$ relates the pseudoscalar meson mass to the quark mass is set as $1.6\pm0.2$ GeV \cite{jhep01010}.

 For the concerned decays, the branching ratios can be written as
 \begin{eqnarray}
\mathcal {B}(B\rightarrow \chi_{c1} K(\pi))=\frac{G_F^2\tau_{B}}{32\pi M}(1-r^2)
|\mathcal {A}|^2.
\end{eqnarray}
Using the above formulas and inputs, we derive the $CP$-averaged  branching ratios for the concerned decays,
 \begin{eqnarray}\label{eq:brs}
\mathcal {B}(B^+\rightarrow \chi_{c1}K^+)&=&(4.4^{+1.4+0.9+0.7+0.2+0.5}_{-1.1-0.8-0.7-0.4-0.4})
\times 10^{-4}=(4.4^{+1.9}_{-1.6})\times 10^{-4},\nonumber\\
\mathcal {B}(B^0\rightarrow \chi_{c1}K^0)&=&(4.1^{+1.3+0.9+0.6+0.2+0.5}_{-1.1-0.8-0.7-0.4-0.4})
\times 10^{-4}=(4.1^{+1.8}_{-1.6})\times 10^{-4},\nonumber\\
\mathcal {B}(B^+\rightarrow \chi_{c1}\pi^+)&=&(1.7^{+0.4+0.4+0.2+0.2+0.1}_{-0.4-0.3-0.2-0.2-0.2})
\times 10^{-5}=(1.7\pm 0.6)\times 10^{-5},\nonumber\\
\mathcal {B}(B^0\rightarrow \chi_{c1}\pi^0)&=&(0.8^{+0.2+0.2+0.1+0.1+0.1}_{-0.2-0.2-0.1-0.1-0.1})
\times 10^{-5}=(0.8\pm 0.3)\times 10^{-5},\nonumber\\
\mathcal {B}(B_s\rightarrow \chi_{c1}\bar{K}^{0})&=&(1.4^{+0.5+0.3+0.2+0.0+0.2}_{-0.4-0.3-0.2-0.1-0.2})
\times 10^{-5}=(1.4\pm 0.6)\times 10^{-5},
\end{eqnarray}
where the second equal-sign in each row denote the central value with all uncertainties added in quadrature.
 There are some theoretical uncertainties in our calculations.
 The first one comes from the nonperturbative parameters $\omega_{b_{(s)}}$
 in $B_{(s)}$ meson wave functions.
 For  $B$ decays, we adopt the value $\omega_b=0.40\pm 0.04$ GeV, which is supported by intensive PQCD
studies \cite{PQCDs}. For $B_s$ meson, we will follow the authors in Ref. \cite{prd76074018} and adopt the value  $\omega_{b_s}=0.50 \pm 0.05$ GeV.
The second error comes from the decay constant of $\chi_{c1}$ meson, which varies $10\%$ for error estimates.
The third error is induced by the chiral scale parameter $m_0=1.6\pm 0.2$ GeV \cite{jhep01010} associated with kaon or pion,
which reflect the uncertainty in the current quark masses.
The fourth   one is from the uncertainty of the heavy quark masses. In the evaluation, we also vary the values of $m_{b,c}$
within a $10\%$ range. The last one is caused by the variation
of the hard scale from $0.75t$ to $1.25t$, which characterizes the size of higher-order corrections to the
hard amplitudes. It is found that
the first three errors are comparable and contribute the main uncertainties in our approach.
While the last scale-dependent uncertainty is less than $15\%$ due to the inclusion of the  vertex corrections.

As noted previously,
many other work have performed a systematic study on the  Cabibbo-favored decays.  
For comparison, we also collect their results in Table \ref{tab:br},
as well as the  current world average values from the PDG \cite{pdg2016}.
The branching ratios of $\mathcal {B}(B^+\rightarrow \chi_{c1}K^+)$ evaluated within LCSR method \cite{plb59191} is
$(5.1\pm 0.5)\times 10^{-4}$, which  match well with our results.
Two earlier papers  \cite{prd87074035, plb568127} also discuss the concerned decays  in the QCDF.
In Ref \cite{prd87074035}, the authors
 treat $\chi_{c1}$ as nonrelativistic bound states and gave
 $\mathcal {B}(B^0\rightarrow \chi_{c1}K^0)=1.79\times 10^{-4}$,
  while in Ref \cite{plb568127},
  where the light-cone wave function is used to describe the $\chi_{c1}$ meson,
  the corresponding value is in the range   $(0.87\sim0.97)\times 10^{-4}$.
   Both  of the two predictions yield much smaller values. 
   However, in   another paper \cite{npb811155}, the authors  revisited the exclusive $B$ decays
   to $P$-wave charmonia in the same framework,  where the   colour-octet contributions are included
   and the charmonium is described as a Coulomb bound state.
   Their theoretical calculations, with reasonable parameter choices,
    can be in qualitative agreement with ours as well as  the experimental data.
  It also can be seen that, for the $K$ decay modes, 
our  calculations in the conventional PQCD scheme  are somewhat larger  than the previous hybrid PQCD ones \cite{prd71114008}
   due to the  different  
   scheme about the  factorizable contributions, the $\chi_{c1}$ DAs,
   and the  vertex corrections as mentioned in the Introduction.

Comparing with the data, our predicted branching ratios of the Cabibbo-favored modes in Eq. (\ref{eq:brs}) 
comply with the world average
$\mathcal {B}(B^+\rightarrow \chi_{c1}K^+)=(4.79\pm0.23)\times 10^{-4}$ \cite{pdg2016} from the measurements \cite{prl96052002,prd66052005,prl102132001,prl107091803}
\begin{eqnarray}
\mathcal {B}(B^+\rightarrow \chi_{c1}K^+)=\left\{
\begin{aligned}
(&4.94 \pm 0.11(\text{stat})\pm 0.33(\text{syst}))\times 10^{-4} \quad\quad\quad  &\text{Belle (2011)}, \nonumber\\ 
(&4.5 \pm 0.1(\text{stat})\pm 0.3(\text{syst}))\times 10^{-4} \quad\quad\quad  & \textit{BABAR}~ (2009),  \nonumber\\ 
(&8.1 \pm 1.4(\text{stat})\pm 0.7(\text{syst}))\times 10^{-4} \quad\quad\quad  & \textit{BABAR}~ (2006),  \nonumber\\ 
(&15.5 \pm 5.4(\text{stat})\pm 2.0(\text{syst}))\times 10^{-4} \quad\quad\quad  &\text{CDF (2002)},  \nonumber\\ 
\end{aligned}\right.
\end{eqnarray}
and $\mathcal {B}(B^0\rightarrow \chi_{c1}K^0)=(3.93\pm0.27)\times 10^{-4}$ from \cite{prd62051101,prl102132001,prl107091803}
\begin{eqnarray}
\mathcal {B}(B^0\rightarrow \chi_{c1}K^0)=\left\{
\begin{aligned}
(&3.78 ^{ +0.17}_{ -0.16}(\text{stat})\pm 0.33(\text{syst}))\times 10^{-4} \quad\quad\quad  &\text{Belle (2011)}, \nonumber\\ 
(&4.2 \pm 0.3(\text{stat})\pm 0.3(\text{syst}))\times 10^{-4} \quad\quad\quad  &\textit{BABAR}~ (2009),  \nonumber\\ 
(&3.1 ^{ +1.6}_{ -1.1}(\text{stat})\pm 0.1(\text{syst}))\times 10^{-4} \quad\quad\quad  &\text{CLEO (2000)}. \nonumber\\ 
\end{aligned}\right.
\end{eqnarray}

\begin{table}
\caption{The branching ratios (in units of $10^{-4}$) of the Cabibbo-favored decays from different theoretical  work \cite{prd71114008,plb59191,plb568127,prd87074035,npb811155}. The data are
 taken from the PDG 2016 \cite{pdg2016}. The original experimental results can be found in \cite{prd62051101,prl96052002,prd66052005,prl102132001,prl107091803}.}
\label{tab:br}
\begin{tabular}[t]{lccccccc}
\hline\hline
Modes  &This Work  & LCSR   \cite{plb59191}&hPQCD \cite{prd71114008} & QCDF-I   \cite{plb568127} & QCDF-II  \cite{npb811155}
 & QCDF-III \cite{prd87074035} &Data \cite{pdg2016} \\ \hline
$B^+ \rightarrow \chi_{c1} K^+ $  &$4.4^{+1.9}_{-1.6} $ & $5.1\pm 0.5$ & $3.15^{+3.17}_{-2.61}$ &-- &--&--&$4.79\pm0.23$ \\
$B^0 \rightarrow \chi_{c1} K^0 $   & $4.1^{+1.8}_{-1.6}$ &-- & $2.94^{+2.97}_{-2.43} $ & $0.87\sim0.97$ & $1.31\sim 10.31$ & $1.79$
&$3.93\pm 0.27$ \\
\hline\hline
\end{tabular}
\end{table}

Now, we turn our attention to  the Cabibbo-suppressed decays. From Eq. (\ref{eq:brs}),
the value of $\mathcal {B}(B_s \rightarrow \chi_{c1} \bar{K}^{0} )$  have a tendency to be
smaller than 2$\mathcal {B}(B^0 \rightarrow \chi_{c1} \pi^{0} )$.
Although  the $B_s$  and $K$ meson decay constants  are larger than those of the $B^0$ and $\pi^0$ meson,
the SU(3) breaking effects in the twist-2 distribution amplitudes of the $K$ meson, parametrized by the first Gegenbauer moment
$a_{1}^K$, gives a negative contribution to the $B_s \rightarrow \chi_{c1} \bar{K}^{0}$ decay,
which  induces the smaller  branching ratio. This is similar to the case of
$B_s \rightarrow \psi(2S) \bar{K}^{*0}$ and $B^0 \rightarrow \psi(2S) \rho^{0}$ decays \cite{epjc77610}.
Experimentally, only the  Belle collaboration reported the results
$\mathcal {B}(B^+\rightarrow \chi_{c1}\pi^+)=(2.4 \pm 0.4(\text{stat})\pm 0.3(\text{syst}))\times 10^{-5}$ \cite{prd74051103}
and $\mathcal {B}(B^+\rightarrow \chi_{c1}\pi^0)=(1.12 \pm 0.25(\text{stat})\pm 0.12(\text{syst}))\times 10^{-5}$ \cite{prd78091104},
which are a little larger than our predictions. 
None the less, taking the errors into consideration,
the theoretical predictions and experimental data  can still  agree with each other.
Since these Cabibbo-suppressed decays are still received less attention 
in other approaches, and we wait for future comparison.

As a cross-check, the ratio of the decay rates for the $B\rightarrow \chi_{c1}\pi$ and $B\rightarrow \chi_{c1}K$
 decays, called $\mathcal {R}_{\pi/K}$ below, can be calculated from
Eq.(\ref{eq:brs}), and are estimated   as
 \begin{eqnarray}
\mathcal {R}_{\pi^+/K^+}=\frac{\mathcal {B}(B\rightarrow \chi_{c1}\pi^+)}{\mathcal {B}(B\rightarrow \chi_{c1}K^+)}=(3.9^{+0.2}_{-0.3})\%,\quad
\mathcal {R}_{\pi^0/K^0}=\frac{\mathcal {B}(B\rightarrow \chi_{c1}\pi^0)}{\mathcal {B}(B\rightarrow \chi_{c1}K^0)}=(2.0^{+0.1}_{-0.2})\%,
\end{eqnarray}
where all uncertainties are added in quadrature.
Because most theoretical uncertainties  are cancelled by the
flavor symmetries in the relative branching ratios,
the total   error of $\mathcal {R}$ are only a few percent, much
smaller than those for the absolute  branching ratios.
As can be seen that  the first ratio is comparable with the Belle measurement \cite{prd74051103},
 \begin{eqnarray}
\mathcal {R}_{\pi^+/K^+}=(4.3\pm0.8\pm0.3)\%.
\end{eqnarray}

Next, we consider the $CP$ asymmetries in these decays.
The direct $CP$ violation for the charged modes,
which arise from the interference between the tree contributions and the penguin contributions,  can be written as
 \begin{eqnarray}
A^{\text{dir}}=\frac{|\mathcal {\bar{A}}|^2-|\mathcal {A}|^2}{|\mathcal {\bar{A}}|^2+|\mathcal {A}|^2},
\end{eqnarray}
where $\mathcal {\bar{A}}$ is the $CP$-conjugate amplitude of $\mathcal {A}$.
In decays of neutral $B$ mesons 
to a final state accessible to both $B$ and $\bar{B}$, the interference between the direct decay and the decay via
oscillation leads to  time-dependent $CP$ asymmetry, which takes the form,
 \begin{eqnarray}
A(t)=-C_f
\text{cos}(\Delta m t)+S_f \text{sin}(\Delta m t),
\end{eqnarray}
where $\Delta m>0$ is the mass difference of the two neutral $B$ meson mass eigenstates.
$S_f$ is referred to as mixing-induced $CP$ asymmetry and $A_f=-C_f$ is the direct $CP$ asymmetry,
which can be expressed as
 \begin{eqnarray}\label{eq:violation}
C_f=\frac{1-|\lambda_f|^2}{1+|\lambda_f|^2},\quad S_{f}=\frac{2\text{Im}(\lambda_f)}{1+|\lambda_f|^2},
\end{eqnarray}
with $\lambda_f=\eta e^{-2i\beta_{(s)}}\frac{\mathcal {\bar{A}}}{\mathcal {A}}$.
$\eta$ is the $CP$ eigenvalue of the final state $f$. $\beta_{(s)}$ is the CKM angle defined as usual \cite{pdg2016}.
 The numerical results for the direct $CP$ asymmetries yield
 \begin{eqnarray}\label{eq:dircp}
A_{\chi_{c1}K}&=&-(1.5^{+0.0+0.0+0.0+0.1+0.4}_{-0.1-0.0-0.1-0.1-0.4})\times 10^{-3}=
-(1.5\pm 0.4)\times 10^{-3},\nonumber\\
A_{\chi_{c1}\pi}&=&(1.1^{+0.0+0.0+0.0+0.2+0.5}_{-0.0-0.0-0.1-0.3-0.5})\times 10^{-2}=
(1.1^{+0.5}_{-0.6})\times 10^{-2}, \nonumber\\
A_{\chi_{c1}\bar{K}}&=&(2.5^{+0.1+0.0+0.0+0.2+0.9}_{-0.1-0.0-0.1-0.3-0.9})\times 10^{-2}=
(2.5\pm 0.9)\times 10^{-2},
\end{eqnarray}
where the errors  induced by the same sources as in Eq. (\ref{eq:brs}).
Unlike the branching ratios, the direct $CP$ asymmetry is not sensitive to the
 nonperturbative parameters related to the initial and final states wave functions, but suffer from large uncertainties due to the hard scale $t$.
Since the charged and neutral decay modes differ only in the lifetimes and isospin factor in our formalism,
they have the same direct $CP$ violations. 
It is found that the direct $CP$ violations are rather small (only $10^{-3}\sim 10^{-2}$) due to the penguin contributions are loop
suppressed with respect to the tree contributions. 
On the experimental side, 
some  direct $CP$ violations  were measured by the Belle collaboration \cite{prd74051103}:
 \begin{eqnarray}\label{eq:dirbelle}
A^{\text{dir}}_{CP}(B^+\rightarrow \chi_{c1}\pi^+)=0.07\pm 0.18\pm0.02,\quad
A^{\text{dir}}_{CP}(B^+\rightarrow \chi_{c1}K^+)=-0.01\pm0.03\pm0.02,
\end{eqnarray}
and $BABAR$ collaboration \cite{prl94141801}:
 \begin{eqnarray}\label{eq:dirbabar}
A^{\text{dir}}_{CP}(B^+\rightarrow \chi_{c1}K^+)=0.003\pm0.076\pm0.017.
\end{eqnarray}
Their weighted average, in fact, enter the numbers given in Eq. (\ref{eq:dirchi1}) are in accordance with our calculations.

Since the neutral final state  $\chi_{c1}K^0$
and its $CP$ conjugate are flavor-specific,
here, we replace it with the $CP$-odd eigenstate  $f=\chi_{c1}K_S$ 
to analyze the  mixing-induced $CP$ asymmetries.
The  obtained results  are listed as:
\begin{eqnarray}\label{eq:mixcp}
S(B^0\rightarrow \chi_{c1}K_S)&=&0.70^{+0.00+0.00+0.00+0.00+0.00}_{-0.00-0.00-0.00-0.00-0.01}=0.70^{+0.0}_{-0.1},\nonumber\\
S(B^0\rightarrow \chi_{c1}\pi^0)&=&-0.62^{+0.01+0.00+0.00+0.01+0.02}_{-0.00-0.00-0.00-0.01-0.01}=-0.62^{+0.02}_{-0.01}, \nonumber\\
S(B_s\rightarrow \chi_{c1}K_S)&=&-0.06^{+0.00+0.00+0.00+0.00+0.01}_{-0.00-0.00-0.00-0.00-0.01}=-0.06\pm 0.01,
\end{eqnarray}
which are less sensitive to the those parameters within their uncertainties.
Experimentally only the first value was direct measured.
The HFAVG \cite{HFLAV2016} quotes $S(B^0\rightarrow \chi_{c1}K_S)=0.632\pm0.099$ \cite{HFLAV2016}
 as the average of the Belle \cite{prl108171802}  and $BABAR$ \cite{prd79072009} data,
which is also  compatible within $1.0\sigma$ 
with  our result in Eq. (\ref{eq:mixcp}). 
In the limit of negligible higher-order contributions,  $S$ can be identified  as $\sin{2\beta}$.
As can be seen,
both   theory and experiment  are close to the current world average value $\sin{2\beta}=0.677\pm 0.020$ \cite{pdg2016},
which suggests that this mode can serve as an alternative place to extract CKM phase $\beta$.
The $B_s\rightarrow \chi_{c1}\bar{K}_S$ decay has not been observed so far. For a similar
$B_s\rightarrow J/\psi \bar{K}_S$ mode, the time-dependent $CP$-violation parameters
have been measured by the LHCb \cite{jhep06131} collaboration,
 \begin{eqnarray}\label{eq:dirbabar}
S(B_s \rightarrow J/\psi K_S)=-0.08\pm0.40\pm0.08.
\end{eqnarray}
The small discrepancy is understandable with respect to the different charmonium states. 
It is hope that the future experiment will provide a direct measurement to the   $B_s\rightarrow \chi_{c1} K_S$ mode.
Our predictions can be used to further explore the properties of the $B_s$ system.

\section{ conclusion}\label{sec:sum}
In the wake of recent measurements of the $P$-wave charmonium productions in the hadronic $B$ decays,
we performed the calculations of the $B_{(s)}\rightarrow \chi_{c1}K(\pi)$ decays by employing the PQCD factorization approach.
The predicted  branching ratios for the $B\rightarrow \chi_{c1}K$ modes are consistent with the data and those from  LCSR method,
while the expectations from the  earlier QCDF  and PQCD 
 are somewhat smaller than the measured values.
Our results for   $\mathcal {B}(B^0\rightarrow \chi_{c1}\pi)$ are smaller than
 those of the Belle measurement
but the discrepancies do not exceed two standard deviations
if one take into account  the experimental uncertainty.
For the $B_s$ modes, the branching ratios amount to the order of $10^{-5}$, letting the
corresponding measurement appear feasible.
We further investigate the measurable $CP$ asymmetries.
The present predictions indicate that
the direct $CP$ asymmetries in these channels are very
small due to the suppressed penguin contributions as we mentioned before.
The mixing-induced $CP$ asymmetry $S(B^0\rightarrow \chi_{c1}K_S)$ is not far away from $\sin{2\beta}$,
and this mode can play an important role in the extraction of the CKM angle $\beta$.
These numbers will be further tested by the LHCb and
Belle-II experiments in the near future.

\begin{acknowledgments}
I would like to acknowledge Ce Meng and Hsiang-nan Li for helpful discussions. This work is supported in
part by the National Natural Science Foundation of China
under Grants No.11605060 and No.11547020, in part
by the Program for the Top Young Innovative Talents of
Higher Learning Institutions of Hebei Educational
Committee under Grant No. BJ2016041, and in part by Training Foundation
of North China University of Science and Technology under Grant
No. GP201520 and No. JP201512.
\end{acknowledgments}

\end{document}